\documentclass[manuscript]{aastex}

\usepackage{color,amsmath,graphicx}

\shorttitle{CME propagation: Where does the solar wind drag take over?}
\shortauthors{ et al.}

\begin{document}


\title{CME propagation: where does aerodynamic drag ``take over''?}


\author{Nishtha Sachdeva\altaffilmark{1} and Prasad Subramanian\altaffilmark{2}}
\affil{Indian Institute of Science Education and Research, Dr. Homi Bhabha Road, Pashan, Pune - 411008, India}
\author{Robin Colaninno\altaffilmark{3}}
\affil{Space Science Division, Naval Reseach Laboratory, Washington, DC 20375, USA}

\author{Angelos Vourlidas\altaffilmark{4}}
\affil{The Johns Hopkins University, Applied Physics Laboratory, Laurel, MD 20723 USA}

% This comes as footnotes

\altaffiltext{1}{nishtha.sachdeva@students.iiserpune.ac.in}
\altaffiltext{2}{also Centre for Excellence in Space Sciences, India, http://www.cessi.in}



\begin{abstract}
We investigate the Sun-Earth dynamics of a set of eight well observed solar coronal mass ejections (CMEs) using data from the STEREO spacecraft.
We seek to quantify the extent to which momentum coupling between these CMEs and the ambient solar wind (i.e., the aerodynamic drag) influences their dynamics.
To this end, we use results from a 3D flux rope model fit to the CME data. We find that solar wind aerodynamic drag adequately accounts for the 
dynamics of the fastest CME in our sample. For the relatively slower CMEs, we find that drag-based models initiated below heliocentric distances ranging from 15 to 50 $R_{\odot}$ 
cannot account for the observed CME trajectories. This is at variance with the general perception that the dynamics of slow CMEs are 
influenced primarily by solar wind drag from a few $R_{\odot}$ onwards. Several slow CMEs propagate at roughly constant speeds above 15--50 $R_{\odot}$. Drag-based models initiated above these heights therefore require negligible aerodynamic drag to explain their observed trajectories.

%{\color{red}Such CMEs experience drag much further out in the heliosphere that is %commonly assumed. }
%It points out that the drag is minimal above a height much larger than the CME onset %height, hinting that the use of constant drag-coefficient values
%from 
%the start of a CME shoud be carried out with caution.}

\end{abstract}

%% example has been keyed in ApJ style. See the instructions to authorsn

\keywords{Sun: Corona --- Sun: Coronal mass ejections (CMEs)---Solar wind}

\section{Introduction}

Coronal Mass Ejections (CMEs) are hot blobs of plasma and magnetic field that occasionally erupt from the solar corona. Some CMEs are directed towards the Earth 
and are primary drivers of the majority of so-called ``space weather'' disturbances that have profound consequences on a variety of technologies that 
we use in everyday life. Space weather predictions are therefore crucially dependent on accurate estimates of the Sun-Earth travel time of Earth-directed CMEs
and their speed upon arrival at the Earth. Although we have some understanding of the factors influencing CME propagation through the inner heliosphere, 
several crucial aspects such as the origin of the driving and drag forces are still somewhat unclear. Needless to say, a thorough understanding of the physical
bases of CME propagation is crucial to enable reliable space weather predictions.

Research on CMEs is often divided into ``initiation'' and ``propagation'' issues. Initiation issues deal with the initial eruption stages of a CME,
while propagation issues concern forces at work during its subsequent propagation through the solar corona and the heliosphere.
The division is somewhat artificial, but one can think of propagation issues as the ones that deal with the post-initiation, ``residual'' acceleration stage of 
CMEs \citep[e.g.,][]{Zha06,Sub07,Gop13}. Based on the temporal evolution
of the kinetic and potential energies of a sample of well observed, plane-of-sky CMEs, \citet{Sub07} showed that they are acted upon by a driving force throughout 
the Large Angle and Spectrometric Coronagaph \citep[LASCO;][]{Bru95} field of view (FOV). Around 17 \% of CMEs in the extensive LASCO dataset are observed to accelerate out to  30 $R_{\odot}$ \citep{StC00}. 
We will concern ourselves here with the propagation stage of CMEs. During the propagation stage, CMEs are thought to be subject to Lorentz self-forces close to 
the Sun ( which may be a continuation of the initiation phase) and to aerodynamic drag due to momentum coupling 
with the ambient solar wind farther out. The interplay between these forces determines the dynamics of a CME as it propagates outwards from the Sun. 
It is generally accepted that: 
\begin{enumerate}
\item
very fast CMEs (generally defined as ones with speeds in excess of 1000 $\rm km\, s^{-1}$ near the Sun) experience all their acceleration 
(due to Lorentz self-forces) very close to the Sun, and are subsequently subjected mostly to deceleration due to aerodynamic drag 
from the ambient solar wind; i.e., they are dragged {\em down} by the solar wind via aerodynamic drag. 
\item
the dynamics of slow CMEs (generally defined as CMEs with speeds comparable to the solar wind speed near the Sun, of the order of a few hundred $\rm km\, s^{-1}$) are governed primarily by solar wind aerodynamic drag. While Lorentz self-forces are expected to play some role,
it is often thought that slow CMEs are primarily dragged {\em up} by the solar wind.
\end{enumerate}

Our understanding of how Lorentz self-forces operate within CMEs is not very robust, although we know that they can accelerate, 
as well as decelerate a CME. While Lorentz self-forces must involve misaligned currents and magnetic fields, it has only recently been 
demonstrated that flux rope CMEs need to be substantially non force-free in order to account for Lorentz self-force driving \citep{Sub14}.
%Early work \citep{Anz79} suggested that it had to
%do with the curvature of magnetic flux ropes. 
There is observational evidence to argue that the magnetic energy contained in flux rope 
CMEs is adequate to account for the mechanical energy required to propagate them from the Sun to the Earth \citep[e.g.,][]{Man04,Man06,Sub07}.
However, the actual nature of Lorentz self-forces has not yet been well understood; some authors incorporate it in the eruption process 
\citep{Ise07,Kli14}, while some treatments \citep[e.g.,][]{Che96} appeal to
poloidal flux injection as a proxy for Lorentz self-forces, and use soft X-ray flare profiles as guides to tailor 
the time profile of poloidal flux injection \citep{Che10}.

On the other hand, the idea of momentum coupling with the ambient solar wind was motivated by the findings of \citet{Gop00}, \citet{Man06}, \citet{Mal09}
and 
others who found that CMEs launched with speeds exceeding that of the solar wind were decelerated in the interplanetary medium, while those launched
with speeds slower than the solar wind were sped up, so that all CMEs tend to equilibrate to the speed of the solar wind.
Using a novel method of identifying an ``average'' CME from LASCO data over a one year period during the approach to solar maximum, \citet{Lew02}
arrived at similar conclusions based on the similarities between the velocity profile of the CME and the ambient solar wind.
Such observations have spurred considerable research on drag-based models (e.g.,   \citealp{Car04};  \citealp{Vrs06}; \citealp{How07}; 
\citealp{Bor09}; \citealp{Byr10}; 
  \citealp{Mal10};\citealp{Vrs10}; \citealp{Tem11}; \citealp{Lug13};
 \citealp{Mis13}; \citealp{Dol14}; \citealp{Vrs13} ; \citealp{Iju14}; \citealp{Tem15}).
Most of these investigations employ an empirical drag coefficient $C_{\rm D}$ 
in order to characterize the interaction between the CME and the solar wind. An exception is \citet{Car04} where $C_{\rm D}$ is motivated from the results of 2.5D MHD
simulations.
\citet{Sub12} (SLB2012 hereafter) have obtained a physical prescription for $C_{\rm D}$ from a calculation of collisionless viscosity in the solar wind. 
The Sun-Earth travel time and near-Earth CME speed predicted by a model using this prescription is found to  agree very well with observations of fast CMEs (SLB2012). 
% Later in this paper, we will also see that the SLB2012 $C_{\rm D}$ prescription also works very well for one of the fastest CMEs ever observed (\citealp{Liu14}; 
% \citealp{Tem15}. {\color{blue} are we keeping this event?}

%In this paper, we determine the heliocentric distance beyond which a 1D CME propagation model that incorporates only %aerodynamic 
%drag decribes its trajectory satisfactorily. We find that this distance depends crucially on the CME initial speed 
%(i.e., whether its a ``slow'' or a ``fast'' CME) in the low corona.
Using a large sample of CMEs observed in LASCO aboard the \textit{SOlar and Heliospheric Observatory mission} \citep[SOHO;][] {Dom95} mission, 
\citet{Mic15} have recently characterized three phases of propagation.
The first phase is the initial acceleration phase (presumably) due to Lorentz self-forces, followed by the second phase where a balance between the Lorentz and drag forces acting on the 
CMEs is achieved. The third phase is when the CME primarily decelerates due to aerodynamic drag at the outer edge of LASCO FOV.
Our approach is similar, but we only seek to differentiate when CMEs are drag-dominated and when they are not. We note that the solar wind drag can serve to decelerate, 
as well as accelerate CMEs.
%For the CMEs in our sample, the height below which no $C_{\rm D}$ model ``works'' is around 15-50 R$_{\odot}$. This also %means that above this height the drag model
%describes the observed trajectory fairly well.}

\section{Data and Models \label{sec2}}

%%%%%%%%%%%%%%%%%% DECIDE if CME 7 is to be kept or removed????????????????????

\subsection{CME Event Sample}
 We use a set of eight extensively studied, Earth-impacting CMEs during the rising phase of solar cycle 24 between March 2010 and June 2011 
\citep{Col12}. These CMEs are observed by the SOHO LASCO, Sun-Earth Connection Coronal and Heliospheric
Investigation \citep[SECCHI;][]{How08} coronagraphs and the Heliospheric Imagers on board the \textit{Solar TErrestrial RElations Observatory mission}
\citep[STEREO;][]{Kai08}. Near-Earth parameters for these CMEs are obtained via in-situ measurements from the WIND spacecraft \citep{Omn}.
The data set is described comprehensively in \citet{Col12} and \citet{Col13}, who have used the Graduated Cylindrical Shell Model (GCS) \citep {The06,The09} to obtain a 3D 
reconstruction of the Sun-Earth trajectory for each of these CMEs. This represents one of the best samples of CMEs for which detailed 3D information 
is available from the Sun to near the Earth. The ninth CME in the dataset of \citet{Col12} involves CME-CME interactions, which cannot be 
accommodated by the simple, drag-based model we employ. We therefore omit that event and study only eight CMEs from the sample.\\
%\textbf{\textit{From this sample, the ninth event has uncommon kinematics which is the result of strong %CME-CME interactions since it is preceded by another
%event. We, therefore, omit this CME event from our study.}}
Table \ref{tbl1} contains salient details for each of these CMEs. The CMEs are hereafter referred to using the serial number assigned in this table.
For each CME,
\citet{Col12} gives the detailed 3D height-time trajectory and flux rope morphology from the Sun to near the Earth.
In addition, we use some other parameters which we describe herein. The quantity $h_{0}$ denotes the height at the first measurement and $h_{f}$ denotes 
the height at
the final measurement. The quantity $v_{0}$ is the initial speed of the CME. We note that CMEs 1, 4, 6 and 8 start out fairly slow, with initial speeds 
$<\,$ 220 km s$^{-1}$, while 
CME 3, 5 and 7 start out relatively faster ($>\,$400 km s$^{-1}$). CME 7 was preceded by another CME, thus the solar wind characteristics are expected to 
be different from the quiescent state.
CME 2 starts out with an initial speed of 916 km  s$^{-1}$ and is 
the fastest event in this sample.

\subsection{Drag Model}
Momentum coupling between the CME and ambient solar wind is typically addressed using the following equation; 
\begin{equation}
 F_{drag}\,=\,m_{cme}\,\frac{dV_{cme}}{dt}\,=-\,\,\frac{1}{2}\,\, C_{\rm D}\,\, A_{cme} \,\,n\,\, m_{p}\, 
 \biggl( V_{cme}-V_{sw} \biggr)\,\,\biggl|V_{cme}-V_{sw}\biggr| \, ,  \label{eq1}
\end{equation}\\
where $F_{drag}$ is the drag force due to momentum coupling between the CME and solar wind, $m_{cme}$ is CME mass,
$V_{cme}$ is CME speed, $C_{\rm D}$ is the 
drag coefficient, $A_{cme}$ is CME area, $n$ is the solar wind density, $m_{p}$ denotes the proton mass and $V_{sw}$ is the ambient solar wind speed.
The dimensionless drag parameter ($C_{\rm D}$) expresses the strength of the momentum coupling between the CME and the solar wind. In fluid dynamics texts 
(e.g., \citealp{Lan87}), an equation of motion 
like Equation~(\ref{eq1}) is used to describe high Reynolds number flows past a solid body, where the bulk of the flow (which can be regarded as a largely potential flow) is 
separated from the body by a turbulent boundary layer. The solid body boundary conditions require that the entire velocity (both the tangential, as well as the normal 
component) vanish at the boundary. The turbulence in the boundary layer (which is rather thin, owing to the large Reynolds number) is 
set up because the velocity needs 
to transition from its bulk value to zero on the boundary over a very short distance; nonlinear terms in the Navier-Stokes equation assume importance and give rise 
to turbulence.

We note that Equation \ref{eq1} is also often expressed in the following form (e.g., \citealp{Car04}; \citealp{Vrs10}):
\begin{equation}
 \frac{F_{drag}}{m_{cme}} \equiv  a_{d} =\,-\,\gamma \,\bigr(V_{cme}-V_{sw}\bigr) \bigl|V_{cme}-V_{sw}\bigr| \, , \label{eq5}
\end{equation}
where, $\gamma \, ({\rm cm}^{-1})$ is given by
\begin{equation}
\gamma\,\,=\,\,C_{\rm D} \frac{n m_{p} A_{cme} }{m_{cme}}  \, .
\label{eq6}
\end{equation}

The cross-sectional area of CME is calculated using $\,A_{cme}(R_{cme})= \pi \, R_{cme}\,W_{cme} $,
where $R_{cme}$ is the minor radius and $W_{cme}$ is the major radius of the elliptical cross section in the x-z plane of a flux-rope CME \citep{The11}.
For each CME in our sample, the GCS fitting yields the major radius, the ratio of the minor to major radius and the half angle which can be used to calculate $R_{cme}$
and $W_{cme}$ using equations 27, 28 and 29 of \citet{The11}.

In this paper we will use the SLB2012 prescription for $C_{\rm D}$, as well as prescriptions where the value of $C_{\rm D}$ is maintained at a 
constant value.
For the sake of completeness, we briefly recapitulate the essentials of the SLB2012 viscosity and $C_{\rm D}$ prescription.
The viscosity ($\nu_{sw}$) in the ambient collisionless solar wind is due to resonant scattering of solar wind protons with an Alfv\'en wave spectrum, 
and can be expressed in analogy with a fluid expression for turbulent viscosity \citep{Ver96}:
\begin{equation}
\nu_{sw} = \sqrt{6}\,\frac{2}{15}\,v_{\rm rms}\,\lambda\, ,
\label{eqvisc}
\end{equation}
where $v_{\rm rms}$ is the rms speed of solar wind protons and $\lambda$ is the mean free path, which we take to be the ion inertial length, 
(\citealp{Lea99, Lea00}; \citealp{Smi01}; \citealp{Bru14}) and is given by 

\begin{equation}
\lambda\,\,=\,\,\frac{v_{a}}{\Omega_{i}} \,\,=\,\,\frac{c}{\omega_{p}} \,\,\sim\,\, 228\,\,n^{-1/2}\,\,k\,m \, .
\label{eqioninertial}
\end{equation}
In Equation~(\ref{eqioninertial}), $v_{a}\,$ is the Alf\'{v}en speed, $\,\Omega_{i}\,$ is  the ion cyclotron frequency, $\,\omega_{p}\,$ is the ion plasma frequency and n is the ambient solar wind 
density.
We note that this prescription is similar to that used by \citet{Col89}; but their prescription is a factor of 3 larger. The treatement of SLB2012 uses the Coles \& Harmon (1989) prescription
for $\lambda$, whereas we take it to  be the ion inertial length.
The CME Reynolds number is defined as
\begin{equation}
Re \equiv \frac{(V_{CME} - V_{sw})\,R_{CME}}{\nu_{sw}} \, ,
\label{eqrey}
\end{equation}
where the quantity $\nu_{sw}$ ($\rm cm^{2}\, s^{-1}$) is the solar wind viscosity calculated using the prescription given in SLB2012.
The drag parameter $C_{\rm D}$ is related to Reynolds number using a fit to the data from \citet{Ach72} for the drag on a sphere at high Reynolds numbers:
\begin{equation}
C_{\rm D} = 0.148 - 4.3 \times 10^4 \, Re^{-1} + 9.8 \times 10^{-9} \, Re \, .
\label{eqcdfit}
\end{equation}

The $C_{\rm D}$ prescription given by Equation~(\ref{eqcdfit}) was experimentally determined for subsonic, high Reynolds number flow past a solid metal sphere (\citealp{Ach72}).
It is worth commenting on its applicability to our situation. Firstly, as noted earlier, the basic drag force law (Equation~\ref{eq1}) that is used extensively in this field is 
really a high Reynolds number, solid body law - two of the assumptions inherent in Equation~(\ref{eqcdfit}) are thus consistent with it. Furthermore, it has been shown that the total (magnetic + particle) pressure exhibits a substantial jump across 
a typical magnetic cloud boundary, suggesting that they are over-pressured structures (\citealp{Rus05}; \citealp{Jia06}) that will not deform in response to tangential stresses,
much like a solid body. Achenbach's drag formula has also been verified by modern detached-eddy simulations of high Reynolds number flows in the super-critical 
regime (e.g., \citealp{Con04}), which is the regime we are interested in. Another aspect to address is the fact that Equation~(\ref{eqcdfit}) is derived from experimental 
results for subsonic flow, whereas the motion of CMEs through the solar wind is certainly supersonic. 
According to the Morokovin hypothesis, which has been verified extensively via numerical simulations (e.g., \citealp{Dua11}) a subsonic turbulent drag law is valid even 
for supersonic flows as long as the fluctuations in the turbulent boundary
layer are incompressible, or subsonic. The turbulent fluctuations in the ambient solar wind are certainly incompressible to a very good approximation 
(e.g., \citealp{Sha10}). This is also evident from the small values of the density fluctuation $\Delta n/n$ in the ambient solar wind 
(e.g., \citealp{Bis14}). The values for the fluctuations in the magnetic field $\Delta B/B$ in the turbulent sheath between the shock and the CME are also as
small as 10\% \citep{Aru13}. Since $\Delta B/B$ is generally representative of $\Delta n/n$ (e.g., \citealp{Spa02}), it follows that the turbulent density 
fluctuations in the sheath region are also fairly incompressible. In summary, the SLB2012 $C_{\rm D}$ 
prescription, as described by Equations~(\ref{eqvisc}), (\ref{eqioninertial}), (\ref{eqrey}) and (\ref{eqcdfit}) is quite appropriate to our situation.

However, as mentioned earlier, we use the SLB2012 $C_{\rm D}$ prescription as well as constant values for $C_{\rm D}$ in the results described below.

\subsection{Models for solar wind density, speed and CME mass}
We next describe the data-based models we use for computing the CME mass ($m_{cme}$), solar wind speed ($V_{sw}$) and solar wind density ($n$) in Equation \ref{eq1}.
The quantity $n_{wind}$ in Table \ref{tbl1} denotes the proton number density at
the Earth, approximately one to two days in advance of the CME and the shock arrival. Assuming that the proton and electron number densities are equal, we use a version of the model given by \citet{Leb98} in order to extrapolate the proton number density Sunwards from the Earth. 
The electron number density at a given heliocentric distance is given by:
\begin{equation}
 n(R)=\biggl ( \frac{n_{wind}}{7.2} \biggr ) \, \biggl [ 3.3 \times 10^{5} R^{-2}+4.1 \times 10^{6} R^{-4}+8 \times 10^{7} R^{-6} \, \biggr ]
 \,\,\,\, {\rm cm}^{-3} \, ,\label{eq2}
\end{equation}
where R is the heliocentric distance in solar radii. The original model of \citet{Leb98} assumes that the number density at 1 AU is 7.2 ${\rm cm}^{-3}$. 
The factor $n_{wind}/7.2$ ensures that the number density given by Equation \ref{eq2} matches the proton number density $(n_{wind})$ measured in-situ by
the WIND spacecraft near the Earth. 
The quantity $n(R)$ thus represents the number density of the ambient solar wind into which a given CME propagates.\\
It is well known that the ambient solar wind speed is an important factor influencing CME propagation (eg., \citealp{Tem11}).
The quantity $v_{wind}$ in Table \ref{tbl1} denotes the near-Earth solar wind speed observed in-situ by the WIND spacecraft, in advance of the CME arrival.
In order to obtain the solar wind speed $V_{sw}(R)$ as a function of heliocentric distance ($R$) along the Sun-Earth line,
we use the solar wind model of \citet{She97,She99}:
\begin{equation}
 V^{2}_{sw}(R)\, = \,v^{2}_{wind}\,\, \bigl[1\,-\,e^{-\frac{(R\,-\,r_{0})}{r_{a}}}\bigr]\, \, .\label{eq3}
\end{equation}
The quantity $r_{0}$, which denotes the heliocentric distance where the solar wind speed is zero, is taken to be 1.5 R$_{\odot}$.
The quantity $r_{a}$, which is the e-folding distance over which the asymptotic speed $v_{wind}$ is reached, is taken to be 50 $R_{\odot}$.
We have verified that our results are not very sensitive to the precise value of $r_{a}$; this will be described later in this paper.

To calculate the CME mass, we used the Thomson scattering geometry from \citet{Bil66}, which relates the observed brightness to coronal electron density.
We calculated the mass in the COR2 FOV assuming that all the mass of the CME is concentrated in a plane at the longitude derived from the GCS model. This assumption 
is a first order estimate of the 3D mass consistent with the method of \citet{Col09}. To correct for the effects of the occulter on the observed 3D mass, we used 
the prescription of \citet{Bei13} who suggest the following functional form in the coronagraphic FOV:
\begin{equation} 
 m_{cme}(R)\,=\,m_{0}+\Delta m\,\,(R\,-\,h_{occ})\, . \, \label{eq4}
\end{equation}
In this equation, $m_{0}$ is the ``true'' mass of the CME when it is first visible above the occulter. 
The quantity $\Delta m$ is the real mass increase with height and $h_{occ}$ is the size (radius) of the occulter in projection (Table \ref{tbl1}). 
These terms in Equation \ref{eq4} are derived from fitting the observed mass increase as the CME emerges from behind the occulter. Thus we 
can estimate the un-occulted 3D mass of the CME in the COR2 FOV with Equation \ref{eq4}. This mass formulation is valid only within the COR2 FOV, 
therefore, 
we assume the mass to be constant beyond 15 R$_{\odot}$. \\
We have also added the virtual mass (e.g. \citealt{Car96}; \citealt{Car04}) to $m_{cme}$ in our calculations. We note that the
addition of the virtual mass does not change the results in any significant manner.

%\subsection{CME Mass Estimation}

\section{Model predictions vs data}
To recapitulate, Equation \ref{eq1} is a simplified, one-dimensional equation of motion describing the dynamics of a CME that is governed
solely by aerodynamic drag due to momentum coupling with the ambient solar wind. Other effects such as Lorentz self-force 
driving are not included in this treatment. Due to momentum coupling between the CME and the solar wind, CMEs that travel slower than the solar wind would thus be accelerated, while those travelling faster than the solar
wind would be decelerated. As described above, we have used realistic observational values for all parameters. The CME speed supplied to the model as an initial condition is estimated from the observed height-time data using
 either a multipolynomial fit as in \citet{Col12} or via numerical differentiation.
We compare the prediction of the CME trajectory from Equation \ref{eq1} with the observed height-time trajectory obtained via 3D GCS 
flux rope fitting for each of the CMEs in Table \ref{tbl1}. Equation \ref{eq1} yields the speed of a CME ($V_{cme}$) as a function of time
from the start of the event.
The output of Equation \ref{eq1} is integrated to obtain a height-time trajectory predicted by the model. We compare the predicted height-time trajectory 
with the observed height-time data for each CME in the next subsection. Other relevant parameters dealing with the comparison between 
the observations and corresponding models are discussed in \S~3.1.

\subsection{Comparision of height-time profiles}

The observed height-time data together with the model predictions are shown in Figures~\ref{fig1} and \ref{fig2}. The results for CMEs 1, 2, 3 and 4 of 
Table~\ref{tbl1} in Figure~\ref{fig1} and those for CMEs 5, 6, 7 and 8 are shown in Figure~\ref{fig2}. 
Observed height-time data points are denoted by diamonds, 
while the lines show the model predictions. The dash-dotted line (red) shows the prediction of the model (Equation~\ref{eq1}) when it is initiated with quantities
corresponding to the first data point. The solid (blue) line, on the other hand, shows the model prediction when it is initiated with these quantities 
corresponding to a later time; in other words, the model CME is initiated from a height larger than the first observation. 
Other details about the model predictions for each CME are discussed in the following sections.

\subsubsection{CME 2}
\begin{figure}
% \epsscale{1.0}
\plotone{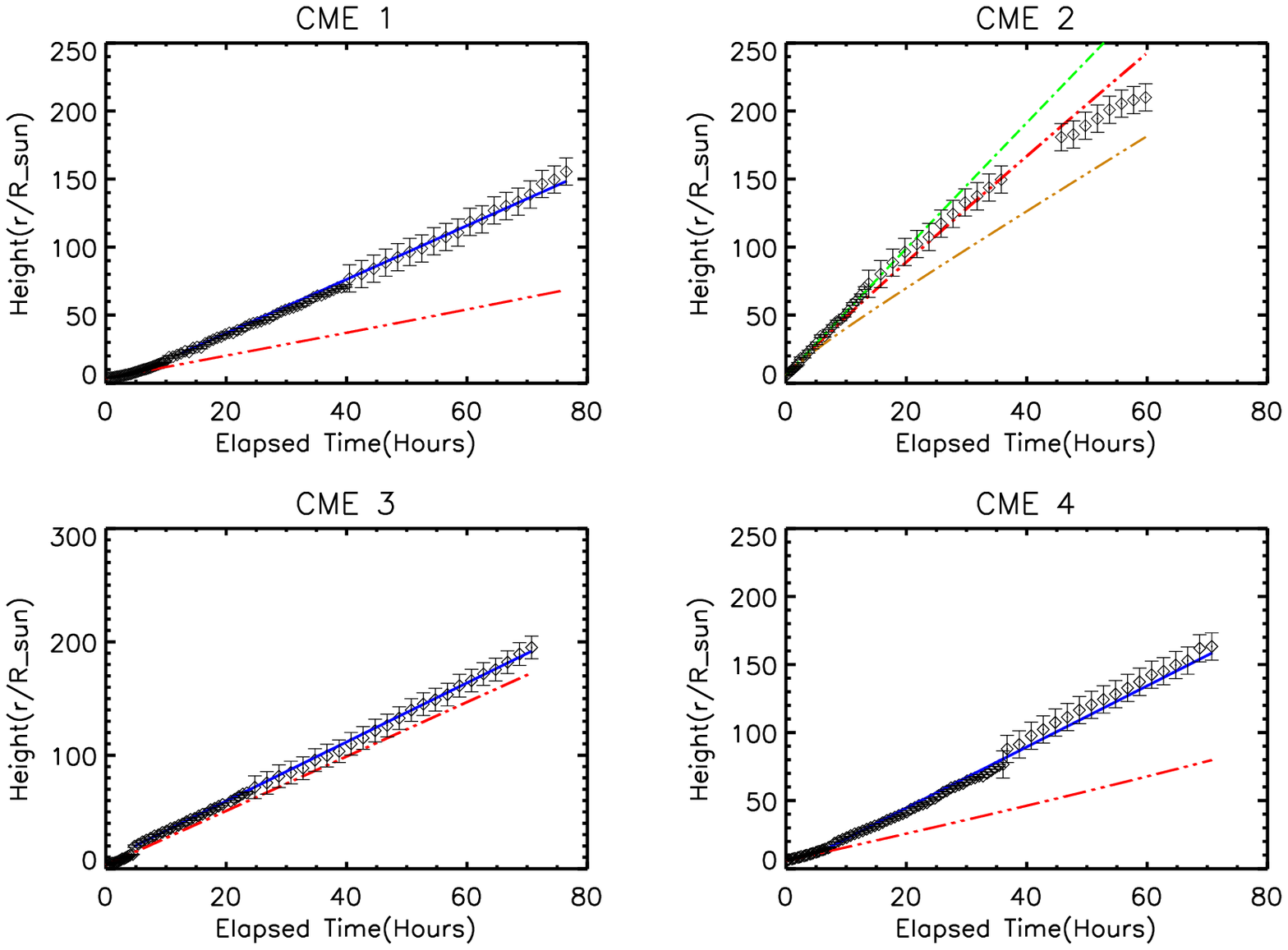}
\caption{Height-time observations and model predictions for CMEs 1, 2, 3 and 4. 
The height-time data points are denoted by diamond symbols along with error bars for COR2 \& HI data. 
The red dash-dotted line denotes the predictions of the model when it is initiated from the first observed data point, while the blue solid line denotes the model 
predictions when it is initiated from $\widetilde{h}_{0}$. In the case of CME 2, the green dash-dotted line indicates the model solutions when initiated from the start with a constant
 $C_{\rm D}$ of 0.1 while the brown dash-dotted line represents the model predictions using a constant $C_{\rm D}$ of 5.0. \label{fig1}}
\end{figure}

\begin{figure}
% \epsscale{1.0}
\plotone{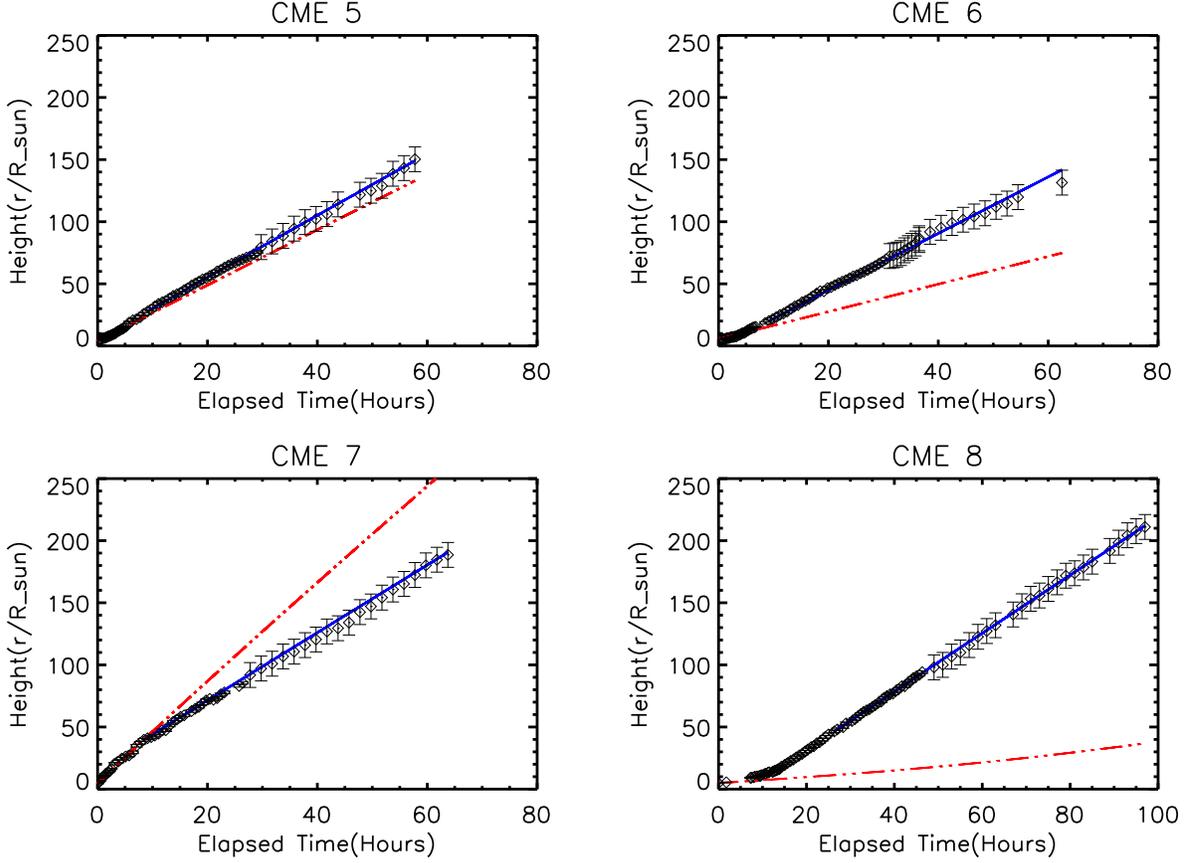}
\caption{Height-time plot for data and derived heights from the drag-only model for CMEs 5, 6 ,7 and 8. The height-time data points are denoted by diamond symbols 
along with error bars for COR2 \& HI data. The dash-dotted lines denoted the predictions of the model when it is initiated from the first observed data point, 
while the solid line denotes the model predictions when it is initiated from $\widetilde{h}_{0}$.\label{fig2}}
\end{figure}

We note that the dash-dotted line which is computed using the SLB2012 $C_{\rm D}$ prescription agrees reasonably with the data points for the 
fastest CME in our sample, CME 2, which has a starting speed 
of 916 $\rm km\, s^{-1}$. 
This is expected, since fast CMEs, which are primarily decelerating, are dominated by solar wind aerodynamic drag, which is the only force
that is included in Equation~\ref{eq1}.\\
% The corresponding height-time profile is shown as the dash-dotted line in Figure \ref{fig1}.
 Although the height-time plot (Figure \ref{fig1}) looks roughly like a straight line, and might suggest that the CME speed is nearly constant, we note that
this is not so; CME 2 decelerates from 
$\sim 916\,\, km\,s^{-1}$ to $\sim 715\,\, km\,s^{-1}$.

The modeled CME is somewhat faster than the observed one; the observed Sun-Earth travel time for 
CME 2 was $\approx$ 60 hours and the model predicts that the CME arrives at the Earth around 8 hours earlier. This represents an error of 14$\%$.
$\Delta \widetilde{h}$ which is the difference between the last measured height from observations and the final modeled height, was -31.1
R$_{\odot}$ for this CME. 

Our time of arrival (ToA) errors are comparable to the values of $\approx$ 10-12 hours obtained
from MHD models (\citealp{May15}) and drag-based models (\citealp{Shi15}).
%{\color{red} Furthermore, \citet{Col09} mention that the heliospheric mass measurements for this CME are probably not %very accurate.}\\
Although we have used observational data as much as possible for the model fitting,
there is some room for uncertainty in the quantity $r_{a}$. We find that a decrease(increase) in $r_{a}$ of 50$\%$ results in a 3.7$\%$ decrease (increase) in the 
predicted CME travel time.
It is also possible that errors in the GCS flux rope fitting procedure can lead to errors in the measured cross-sectional area of the CME.
We find that a 50 \% decrease in the CME area results in a {\bf 6 \% }decrease in the CME travel time.\\
The range of C$_{\rm D}$ values predicted by SLB2012 $C_{\rm D}$ prescription for CME 2 is 0.6--1.4. Model predictions using a constant $0.6\,<\,C_{\rm D}\,<\,2.0$ are 
reasonably similar to the dash-dotted (red) line of CME 2 in Figure \ref{fig1} which uses 
the SLB2012 $C_{\rm D}$. On the other hand, constant $C_{\rm D}$ models with values for $C_{\rm D}$ that lie substantially outside of $0.6\,<\,C_{\rm D}\,<\,2.0$ perform poorly.
This is evident from the examples shown in Figure \ref{fig1}; model solutions with constant $C_{\rm D}$s of 0.1 (green dash-dotted line) and 5 (brown dash dotted line) 
disagree considerably with the observed height-time data for CME 2.
For the $C_{\rm D}$ = 0.1 model, the difference between the 
 predicted and the observed ToA at 180 R$_{\odot}$ is $\approx$ -8 hours. For $C_{\rm D}$ = 5.0, it is $\approx$ 14 hours, and only -1.8 hours for 
 the SLB2012 $C_{\rm D}$ prescription.

% {\color{red} All bold removed above here}

\subsubsection{Slower CMEs}
Next, we turn our attention to the somewhat slower CMEs in our sample (Table \ref{tbl1}): CMEs 1, 3, 4, 5, 6, 7 and 8. 
For such CMEs, it is generally believed that Lorentz self-forces are dominant for the initial part of its trajectory, 
 while aerodynamic drag takes over at larger heights (e.g., \citealp{Mic15}). Some authors (e.g., \citealp{Lew02}) claim that representative slow CMEs are exclusively dragged {\em up} by the solar wind. Some authors (e.g., \citealp{Mis13}) apply the drag-based model 
to explain observations of slow CMEs over its entire trajectory while some authors (e.g., \citealp{Byr10,Car12,Gop13}) 
think that solar wind aerodynamic drag becomes dominant
for slow CMEs beyond a few solar radii. 
%All the aforementioned studies employ an ad-hoc constant $C_{\rm D}$.
It is evident from Figures \ref{fig1} \& \ref{fig2} that the dash-dotted line disagrees considerably with the data (denoted by diamonds) for all events except
CME 2.
The maximum height (at the last timestamp) predicted by the model 
(when initiated from  the start) falls short of the final observed height ($h_{f}$) for each CME (except CME 7, in which case it is much 
larger than $h_{f}$). We find that the difference between the final 
height predicted by the model and the observed final height is large, ranging between 17--175 R$_{\odot}$ for CMEs 1, 3, 4, 5, 6, 7 \& 8. 
This suggests that there is a large discrepancy in the predicted and observed trajectory when the model is initiated from low heights. 
In other words, momentum coupling with the ambient solar wind alone does not satisfactorily explain the dynamics of the CMEs other than CME 2. 
%{\bf suggesting that force due to solar wind coupling only is inadequate in 
%explaining the observed dynamics from the CME onset heights.}

While the dash-dotted (red) lines use the SLB2012 $C_{\rm D}$ prescription, we also find that constant $C_{\rm D}$ 
models with $0.1 < C_{\rm D} < 5$ cannot account for the data when initiated from the first observation. 
This points to the conclusion that CMEs 1, 3, 4, 5, 6, 7 and 8 cannot be considered to be drag-dominated (either drag-accelerated or decelerated) 
from the start. 
This conclusion is independent of the specific $C_{\rm D}$ model 
used (the SLB2012 $C_{\rm D}$ or a constant $C_{\rm D}$).
In order to find the height where the solar wind drag ``takes over'' for the relatively slower CMEs we adopt the following strategy  - 
we initiate the model at a larger height. \\
The initial conditions supplied to the differential equation (Equation \ref{eq1}), such as the CME 
initial speed, mass, cross-sectional area, solar wind speed, background density are those appropriate to this height, which is larger than that at the 
first observed data point.
We repeat this exercise with increasing initiation heights until there is a reasonable agreement between the modeled (solid line) and observed
height-time data.
We carry out this exercise using the SLB2012 $C_{\rm D}$ model as well as constant $C_{\rm D}$ models. As with CME 2, we find that the values for $C_{\rm D}$
predicted by the SLB2012 prescription are a good guide for the constant $C_{\rm D}$ model. In other words, only those constant $C_{\rm D}$ models with $C_{\rm D}$ 
values that are close to those predicted by SLB2012 prescription (last column in Table~\ref{tbl2}) agree reasonably well with the data.
The initiation height beyond which the solid line agrees reasonably with the data is denoted by $\widetilde{h}_{0}$ in Table~\ref{tbl2}, 
and the CME speed at $\widetilde{h}_{0}$ is denoted by $\widetilde{v}_{0}$.
At heights below $\widetilde{h}_{0}$, no drag model yields good agreement with observations. We have verified this by initiating the model at different heights ranging from 
the first observed height ($h_{0}$) to $\widetilde{h}_{0}$, using the SLB2012 $C_{\rm D}$ prescription, as well as the constant $C_{\rm D}$ prescription. Since we do not include any forces other than solar wind 
drag, the only definitive statement we can make about CMEs 1, 3, 4, 5, 6, 7 and 8 is that they are not drag dominated when the model is initiated at {\em any height 
below $\widetilde{h}_{0}$}.

%\subsection{Model predictions and comparison with observations}
Table \ref{tbl2} also lists for each CME the quantity 
$\widetilde{h}_{f}$ which is the final height predicted by the model when it is initiated from $\widetilde{h}_{0}$ (corresponding to the solid lines in 
Figures \ref{fig1} and \ref{fig2} and dash-dotted line for CME 2). $\Delta \widetilde{h}$ is the difference between $h_{f}$ and $\widetilde{h}_{f}$. 
% The quantity $\overline{h}_{f}$ in Table \ref{tbl2} is the final height achieved by the model when initiated from the first observed timestamp. 
% Difference between $h_{f}$ and $\overline{h}_{f}$ is denoted by $\Delta \overline{h}$. 
%{\color{red} It can be easily seen from Figures \ref{fig1} and \ref{fig2} that the model when initiated from the start %reaches a final height very different
%from the observed height as compared to when it is initiated from $\widetilde{h}_{0}$, in which case, $\Delta %\widetilde{h}$ is small for all CMEs.}

%Our results for all CMEs (except CME 2) thus seem to suggest that the solar wind drag %provides a satisfactory description of their dynamics above $\widetilde{h}_{0}$. 
%This conclusion holds for the SLB2012 $C_{\rm D}$ models as well as constant $C_{\rm D}%$ ones ({\bf $C_{\rm D}$ values close to those predicted by 
%SLB2012.) As mentioned before, below the height $\widetilde{h}_{0}$, for any value of %the drag coefficient we find that there is a significant deviation 
%from the observations.}

For CMEs 1, 3, 4, 5 and 6, we have reconstructed 3D data only up to $\sim$ 160 $R_{\odot}$. In order to avoid empirical extrapolation, we therefore compare the 
model velocity predictions with the observed quantities at the last timestamp for all CMEs.
Table \ref{tbl4} lists the quantities $V_{ldp}$, which is the velocity observed at the last timestamp (at $h_{f}$) for each event and $V_{pred}$, which is the 
predicted velocity at the same instant. 
$\Delta V$ is the difference between $V_{pred}$ and $V_{ldp}$. A positive $\Delta V$ means that the predicted CME velocity at last data point
is larger than the observed one, while a negative $\Delta V$ means the converse. 
%Since data for most of the CME extends only upto 130-160 R$_{\odot}$, we make 
%velocity comparisions at the last observed timestamp. 
The ratio of $\Delta V$ to the observed velocity $V_{ldp}$ (at heliocentric height $h_{f}$) is also given in Table \ref{tbl4}. 
$\Delta V/V_{ldp}$ ranges between 0.3$\%$ to 31$\%$ which indicates that the model performs well. Depending on how the ambient solar wind speed compares with 
$\widetilde{v}_{0}$, the solar wind drag decelerates/accelerates the CMEs. 

As with the velocity comparisons, in order to avoid empirical extrapolation, we compare the Time of Arrival (ToA) for each event at two heights: 
130 R$_{\odot}$ and 180 R$_{\odot}$. Only CMEs 2, 3, 7 and 8 attain heights $>$ 180 R$_{\odot}$; therefore ToA comparisions 
at 180 R$_{\odot}$ are made for only these four events. For the remaining events we make ToA comparisons at 130 R$_{\odot}$.
Table \ref{tbl4} describes the quantity $\Delta T$ which is the difference between the predicted and observed ToA at these heights, when the model 
is initiated at $\widetilde{h}_{0}$ (except for CME2, where the model is initiated at $h_{0}$).
%For each representative height, 130 R$_{\odot}$ adn 180 R$_{\odot}$ the predicted ToA is calculated from the solution to %the drag-based model when it is initiated 
%from $\widetilde{h}_{0}$ (except for CME 2, in which case it is the same as $h_{0}$). 
A positive $\Delta T$ means that the arrival time of the model CME is delayed with respect to the observed one, while a negative $\Delta T$ denotes its converse.
 The quantity $\Delta T/T$ denotes the ratio of $\Delta T$ to the observed CME transit time up to each height. 
 It ranges between 1$\%$ and 5$\%$ for 130 R$_{\odot}$ and for 180 R$_{\odot}$ it ranges between 0.1$\%$ and 4$\%$. %Thus, $\Delta T/ T$ is remarkbly small 
%for all CMEs at both these representative heights.

An important point to be noted is that that most of the slow CMEs show little evolution in their speeds beyond $\widetilde{h}_{0}$.
While CME 5 decelerates moderately ($\sim$ 13 $km\,\,s^{-1}$) between 28 $R_{\odot}$ and 150 $R_{\odot}$, CME 1 
hardly decelerates. As before, constant $C_{\rm D}$ models with values of $C_{\rm D}$ that are close to the SLB2012 $C_{\rm D}$ predictions 
yield similar results. Since the deceleration is so small, even constant $C_{\rm D}$ models with $C_{\rm D} \approx 0$ yield good results.
% While the quantity $\Delta T/T$ is quite insensitive to the value of $C_{\rm D}$ used in constant $C_{\rm D}$ models, 
% this is not the case with $\Delta V/V_{ldp}$. 
% {\color{red} For example, for CME 1 with $C_{\rm D}$=0, there is negligible change in both $\Delta T/T$ and
% $\Delta V/V$. However, in case of CME 5, $\Delta T/T$ and
% $\Delta V/V$ decreases by 2 \% and increases by 1.2 \% respectively.}
Initiating drag-based models above $\widetilde{h}_{0}$ thus does not constrain the drag-based models to any appreciable extent,
and only reinforces the fact that most slow CMEs don't accelerate or decelerate much beyond $\widetilde{h}_{0}$.

\section{Discussion and Conclusions}

We worked with a dataset of eight well observed CMEs that were tracked by the SOHO LASCO and SECCHI coronagraphs and the HI imagers aboard the STEREO spacecraft \citep{Col12}. 
These events were fitted with the GCS model, which yields the 3D height-time profile of the CME as it propagates through the heliosphere (Table~\ref{tbl1}). 
Despite the relatively small number of events, our sample covers a wide range of speeds: some events start out 
fairly slow (around a few hundred $\rm km\, s^{-1}$), some have intermediate speeds and one is relatively fast (starting speed of 916 $\rm km\, s^{-1}$), although it is certainly not among the fastest CMEs observed. 
%While it is generally accepted that CMEs are subject to Lorentz self-forces and are also regulated (via momentum coupling) by the ambient solar wind, 
%the relative importance of these effects is not very well understood. 
We focused here on the effect of solar wind aerodynamic drag on CMEs using a simple, widely used 1D model (Equation \ref{eq1}) that allows one to obtain the
predicted height-time trajectory of a CME that is subject only to this drag (Lorentz self-forces are not included). It may be noted that solar wind 
drag can serve to decelerate, as well as accelerate a CME. We used two prescriptions for the drag parameter $C_{\rm D}$: the SLB2012 prescription, and a 
prescription where the value of $C_{\rm D}$ is maintained at a constant value. For a given CME, we use observationally determined values 
for the CME cross-sectional area, mass, initial speed, ambient solar wind density and speed to solve Equation \ref{eq1} and obtain the predicted height-time
trajectory, which we compare with the observed one. 
In general, such a model is expected to work best for very fast CMEs, for which (accelerating) Lorentz self-forces are expected to cease very soon after 
initiation (e.g., \citealp{Che10}), and the only operating force from then onwards is the (decelerating) solar wind aerodynamic drag. In keeping with this expectation, 
we find that the drag-based model performs well in describing the dynamics of the fastest CME in our sample, CME 2.

Its generally thought that slow CMEs are ``picked up by the solar wind'' (i.e., accelerated by aerodynamic drag) from as early 
as a few $R_{\odot}$. Results from drag-based models reported in the most of the existing literature refer to initiation from the first observed timestamp,
which typically correspond to only a few $R_{\odot}$.
For the slow CMEs in our sample, we find that the predictions of drag-based models do not 
agree with the observations when initiated at any heliocentric height
below $\widetilde{h}_{0}$ ($\widetilde{h}_{0}$ for slow CMEs ranges from 15--50 $R_{\odot}$). When initiated at 
heights $< \widetilde{h}_{0}$, the errors in predicting the final height for slow CMEs range from 12 \% to 80 \%.
Our results thus suggest that solar wind drag is not the dominant influence on CME trajectories until 15--50 $R_{\odot}$. 
It might be important to include other forces such as Lorentz forces, at least as far as 15 to 50 $R_{\odot}$. 
Recent work by \citet{Zic15} also assumes that solar wind drag is effective only above 15 R$_{\odot}$.

The speeds of several slow CMEs in our sample stay roughly constant for heights $ > \widetilde{h}_{0}$. Although drag-based 
models initiated above $\widetilde{h}_{0}$ seem to generally perform well, solar wind drag in fact ``doesn't have much to do'' for these CMEs 
(since the CME speed evolution at heights $ > \widetilde{h}_{0}$ is negligible) and the drag parameter $C_{\rm D}$ is not well constrained.

Acknowledgement:
NS acknowledges support from a PhD studentship at IISER Pune. PS acknowledges support from the Asian Office of Aerospace Research and Development. RC and AV are supported by NASA contract S-136361-Y to NRL. The SECCHI data are produced by an international consortium of the NRL, LMSAL and NASA GSFC (USA), RAL and 
Univ. Bham (UK), MPS (Germany), CSL (Belgium), IOTA and IAS (France). We acknowledge a very thorough review by the referee, which helped us considerably in improving this 
paper.

% {\bf In table 2, They have asked to remove $\Delta \overline{h}$, however we have described it in our text.}
%{\bf \textit{Note \\ Sir, the enteries for CME 5 are using values 10 \% more than the polynomial fitted value %of 520 $\rm km\, s^{-1}$.\\ Word Count: \\
%Abstract: 4 words extra\\ Remaining text: approx 700 words extra.}}\\

%BIBLIOGRAPHY(Alphabetical)

% figures

% 
% \begin{figure}
% % \epsscale{1.0}
% \plotone{cme2_CD.ps}
% \caption{Height-time plot for data and derived heights from the drag-only model for CME 2. The height-time data points are denoted by diamond symbols 
% along with error bars for COR2 \& HI data. The dash-dotted lines denote model predictions when it is initiated from the first observed data point with different 
% $C_{\rm D}$ models. Red dash dotted line is the solution to drag model using the SLB2012 $C_{\rm D}$ prescription. 
% The green dash-dotted line represents model solutions with $C_{\rm D}$ = 0.1 and brown dash-dotted line is the solution for $C_{\rm D}$ = 5.0.
% \label{fig3}}
% \end{figure}

%%%%%%%%%%%% TABLE 1

\clearpage
\begin{deluxetable}{clcccccccc}
\tabletypesize{\scriptsize}
\tablecaption{Observed CME and solar wind parameters\label{tbl1}}
\tablewidth{0pt}
\tablehead{
\colhead{No.} & \colhead{Date} & \colhead{$h_{0}$} &  \colhead{$h_{f}$}
&\colhead{$v_{0}$} & \colhead{$n_{wind}$} & \colhead{$v_{wind}$}  
& \colhead{$log (m_{0})$} & \colhead{$h_{occ}$} &  \colhead{$log(\Delta m)$} \\
\colhead{} & \colhead{} & \colhead{(R$_{\odot}$)}& \colhead{(R$_{\odot}$)}& \colhead{($\rm km\,s^{-1}$)}& \colhead{($\rm cm^{-3}$)}& \colhead{($\rm km\,s^{-1}$)}
&\colhead{} &\colhead{(R$_{\odot}$)} &\colhead{}}
\startdata
1 & 2010 Mar 19 - 2010 Mar 23   & 3.5  & 155.5 & 162  & 3.60 & 380  & 14.7 & 3.65 & 14.3 \\
2 & 2010 Apr 03  - 2010 Apr 05   & 5.5  & 210.0 & 916  & 7.15 & 470 & 15.1 & 3.46 & 14.6 \\
3 & 2010 Apr 08 - 2010 Apr 11    & 2.9  & 195.1 & 468  & 3.60 & 440 & 15.4 & 3.55 & 14.7 \\
4 & 2010 Jun 16 - 2010 Jun 20    & 5.7  & 163.3 & 193  & 3.50 & 500 & 14.9 & 4.37 & 14.1\\
5 &2010 Sep 11  - 2010 Sep 14  & 4.0  & 150.3 & 444  & 4.00 & 320  & 15.2 & 3.84 & 14.7\\
6 & 2010 Oct 26 - 2010 Oct 31   & 5.3  & 131.5 & 215  & 3.80 & 350 & 15.2 & 5.66 & 14.7\\
7 & 2011 Feb 15 - 2011 Feb 18   & 4.4  & 188.6 & 832  & 2.50 & 440  & 15.6 & 3.85 & 13.9\\
8 & 2011 Mar 25 - 2011 Mar 29    & 4.8  & 211.0 & 47  & 3.00 & 360  & 15.4 & 4.43 & 14.3\\
%9 & 20110601-20110604 & 5.00 & 4.571 & \nodata & 210.000 & \nodata & 390  \\

\enddata
\tablecomments{Column 1: Serial number; Column 2: Observed date range for the event; Column 3: Initial observed height;
Column 4: Final measured height; Column 5: Initial CME speed; Column 6: Observed proton number density at 1 AU;
Column 7: Observed solar wind speed at 1 AU; Column 8: log of the ``true'' CME mass at the first measurement; 
 Column 9: Occulation height; Column 10: log of mass increase per height}
\end{deluxetable}

\begin{deluxetable}{ccccccc}
%\tabletypesize{\scriptsize}
\tablecaption{CME model parameters and SLB2012 $C_{\rm D}$ \label{tbl2}}
%  \tabletypesize{\scriptsize}
\tablewidth{0pt}
\tablehead{
\colhead{No.} & \colhead{$\widetilde{h}_{0}$} & \colhead{$\widetilde{v}_{0}$}  &
\colhead{$\widetilde{h}_{f}$} &  \colhead{$\Delta \widetilde{h}$} & \colhead{SLB2012 $C_{\rm D}$} \\
\colhead{} & \colhead{(R$_{\odot}$)} & \colhead{($\rm km\,s^{-1}$)} &\colhead{(R$_{\odot}$)} &\colhead{(R$_{\odot}$)} & \colhead{}}
\startdata
1 &   21.9 &  383  & 148.3 &   7.1  & 0.1--0.3\\
2 &   5.5  &  916  & 241.1 & -31.1 &  0.6--1.4\\
3 &   19.7 &  506  & 191.8 &   3.3  &  0.2--0.4\\
 4 &   15.2 &  437  & 158.4 &  4.8  &  0.2--0.3\\
5 &   27.7 &  490  & 149.1 &   1.2  &  0.4--0.6\\
6 &   20.1 &  445  & 141.7 &  -10.3 &  0.25--0.5\\
 7 &   39.7 &  530  & 190.7 &   -2.1  &  0.25--0.4\\
8 &   46.5 &  456  & 212.0 &  -1.0  &  0.25--0.35\\
%9 & \nodata&  \nodata & \nodata &  \nodata & \nodata \\
\enddata

\tablecomments{Column 1: CME number corresponding to Table~\ref{tbl1};
Column 2: Height at which model is initiated; Column 3: CME speed at height $\widetilde{h}_{0}$; 
Column 4: Final height at last 
observed timestamp when the model is initiated from $\widetilde{h}_{0}$; Column 5: Difference between $h_{f}$ and $\widetilde{h}_{f}$; Column 6: 
Range of $C_{\rm D}$ values predicted by SLB2012 $C_{\rm D}$
prescription.}

\end{deluxetable}
%Time of arrival 

\begin{deluxetable}{ccccccccc}
\tablecaption{CME arrival time and speeds : observations vs predictions \label{tbl4}}
% \tabletypesize{\scriptsize}
\tablewidth{0pt}
\tablehead{ \multicolumn{5}{c}{} & \multicolumn{2}{c}{130 R$_{\odot}$} & \multicolumn{2}{c}{180 R$_{\odot}$} \\
\colhead{No.} & \colhead{$V_{ldp}$} & \colhead{$V_{pred}$} & \colhead{$\Delta V$} & \colhead{$\Delta V/V_{ldp}$} &  \colhead{$\Delta T$} & \colhead{$\Delta T/T$} &
\colhead{$\Delta T$} & \colhead{$\Delta T/T$}\\
\colhead{} & \colhead{($km \,\,s^{-1}$)} & \colhead{($km \,\,s^{-1}$)} & \colhead{($km \,\,s^{-1}$)} & \colhead{$\%$}  & \colhead{hr} & \colhead{$\%$} &  \colhead{hr} 
& \colhead{$\%$}}
\startdata
1 & 439 & 382 & -57 & 13 & 0.85 & 1 & \nodata & \nodata \\
2 & 639 & 731 & 92 & 14 & \nodata & \nodata & -1.8 & 4 \\
3 & 501 & 503 & 2 & 0.3 & \nodata & \nodata & 0.07 & 0.09 \\
4 & 501 & 483 & -18 & 3 & 2.5 & 4.5 & \nodata  & \nodata \\ 
5 & 564 & 477 & -87 & 15 & -1.98 & 3.8 & \nodata & \nodata  \\
6 & 457 & 439 & -18 & 4 &  -0.75 & 1.27 & \nodata & \nodata \\
7 & 595 & 526 & -69 & 12 & \nodata & \nodata & 0.12 & 0.2\\
8 & 342 & 451 & 109 & 31 & \nodata & \nodata & -0.39 & 0.46\\
\enddata
\tablecomments{Column 1: CME number correspoing to Table \ref{tbl1}; Column 2: Observed CME velocity at last observed timestamp;
Column 3: Predicted model velocity at the last timestamp when the model is initiated from $\widetilde{h}_{0}$ (except CME 2 in which case 
 model starts from $h_{0}$); Column 4: $\Delta V$ is the difference between $V_{pred}$ and $V_{ldp}$; Column 5: Ratio of $\Delta V$ and $V_{ldp}$;
Column 6: Difference between the Predicted and Observed ToA at 130 R$_{\odot}$; Column 7: Ratio of $\Delta T$ and the observed ToA at 130 R$_{\odot}$;
Column 8: Difference between the Predicted and Observed ToA at 180 R$_{\odot}$; Column 9: Ratio of $\Delta T$ and the observed ToA at 180 R$_{\odot}$.} 
\end{deluxetable}

\begin{thebibliography}{}
\bibitem[Achenbach(1972)]{Ach72} Achenbach, E. 1972, J. Fluid Mech., 54, 565
\bibitem[Arunbabu et al.(2013)]{Aru13} Arunbabu, K. P., Antia, H. M., Dugad, S. R.,et al. 2013, Astronomy \& Astrophysics, 555, A139
\bibitem[Bein et al.(2013)]{Bei13}Bein, B. M., Temmer, M., Vourlidas, A.,et al. 2013, \apj, 768, 31 
\bibitem[Billings(1966)]{Bil66} Billings, D. E. 1966, A guide to the solar corona, (New York: Academic Press)
\bibitem[Bisoi et al.(2014)]{Bis14} Bisoi, S. K., Janardhan, P., Ingale, M.,et al. 2014, \apj, 795, 69
\bibitem[Borgazzi et al.(2009)]{Bor09}Borgazzi, A., Lara, A., Echer, E.,et al. 2009, A \& A, 498, 885
\bibitem[Brueckner et al.(1995)]{Bru95} Brueckner, G. E., Howard, R.A., Koomen, M. J.,et al. 1995, \solphys, 162, 357
\bibitem[Bruno \& Trenchi(2014)]{Bru14} Bruno, R., \& Trenchi, L. 2014, \apj, 787, L24
\bibitem[Byrne et al.(2010)]{Byr10} Byrne, J. P., Maloney, S. A., McAteer, R. T. J.,et al. 2010, \nat \,Communications, 1, 74 
\bibitem[Cargill et al.(1996)]{Car96} Cargill, P. J., Chen, J., Spicer, D. S., \& Zalesak, S. T. 1996, \jgr, 101, 4855
\bibitem[Cargill(2004)]{Car04} Cargill, P. J. 2004, \solphys, 221, 135
\bibitem[Carley, McAteer \& Gallagher (2012)]{Car12} Carley, E. P., McAteer, R. T. James, Gallagher, P. T. 2012, \apj, 752, 36
\bibitem[Chen(1996)]{Che96} Chen, J. 1996, \jgr, 101, 27499
\bibitem[Chen \& Kunkel(2010)]{Che10} Chen, J., \& Kunkel, V. 2010, \apj, 717, 1105
\bibitem[Colaninno \& Vourlidas(2009)]{Col09} Colaninno, R. C., \& Vourlidas, A. 2009, \apj, 652, 1747
\bibitem[Colaninno(2012)]{Col12} Colaninno, R. C. 2012, PhD thesis, George Mason University, arXiv:1206.4290
\bibitem[Colaninno, Vourlidas \& Wu(2013)]{Col13} Colaninno, R. C., Vourlidas, A., \& Wu, C. C. 2013, \jgr, 118, 6866
\bibitem[Coles \& Harmon(1989)]{Col89} Coles, W. A., \& Harmon, J. K. 1989, \apj, 337, 1023
\bibitem[Constantinescu \& Squires(2004)]{Con04} Constantinescu, G., \& Squires, K. 2004, Physics of Fluids, 16, 1449
\bibitem[Dolei et al.(2014)]{Dol14} Dolei, S., Bemporad, A., \& Spadaro, D. 2014, A \& A, 562, A74
\bibitem[Domingo, Fleck \& Poland(1995)]{Dom95} Domingo, V., Fleck, B., Poland A. I. 1995, \solphys, 162, 1
\bibitem[Duan, Beekman \& Martin(2011)]{Dua11} Duan, L., Beekman, I., \& Martin, M. P. 2011, J. Fluid Mech., 672, 245
%\bibitem[Forbes(2000)]{For00} Forbes, T. G. 2000, \jgr, 105, 23,153
\bibitem[Gopalswamy et al.(2000)]{Gop00} Gopalswamy, N., Lara, A., Lepping, R. P.,et al. 2000, \grl, 27, 145
\bibitem[Gopalswamy(2013)]{Gop13} Gopalswamy, N. 2013, in ASI Conf. Ser. 10, STEREO and SOHO contributions to coronal mass ejection studies: some recent results,
ed. N. Gopalswamy, S. S. Hasan, P. B. Rao \& P. Subramanian (Pune, India:ASI), 11 
\bibitem[Howard et al.(2007)]{How07} Howard, T. A., Fry, C. D., Johnston, J. C.,et al. 2007, \apj, 667, 610
\bibitem[Howard et al.(2008)]{How08} Howard, R. A., Moses, J. D., Vourlidas, A.,et al. 2008, \ssr, 136, 67
\bibitem[Iju,Tokumaru \& Fujiki(2014)]{Iju14} Iju, T., Tokumaru, M., \& Fujiki, K. 2014, \solphys, 289, 2157
\bibitem[Isenberg \& Forbes(2007)]{Ise07} Isenberg, P.  A., Forbes, T. G. 2007, \apj, 670,1453
\bibitem[Jian et al.(2006)]{Jia06} Jian, L., Russell, C. T., Luhmann, J. G., Skoug, R. M. 2006, \solphys, 239, 393
\bibitem[Kaiser et al.(2008)]{Kai08} Kaiser, M. L., Kucera, T. A., Davilla, J. M.,et al. 2008, \ssr, 279, 477
\bibitem[Kliem et al.(2014)]{Kli14} Kliem, B., Lin, J., Forbes, T. G.,et al. 2014, \apj, 789,46
\bibitem[Landau \& Lifshitz(1987)]{Lan87} Landau, L. D., \& Lifshitz, E. M. 1987, Fluid Mechanics (2nd ed.; Pergamon Press)
% \bibitem[Lara et al.(2011)]{Lar11} Lara, A., Flandes, A., Borgazzi, A., \& Subramanian, P. 2011, \jgr, 116, A12102
\bibitem[Leamon et al.(1999)]{Lea99} Leamon, R. J., Smith, C. W., Ness, N. F.,et al. 1999, \jgr, 104, 22331
\bibitem[Leamon et al.(2000)]{Lea00} Leamon, R. J., Matthaeus, W. H., Smith, C. W.,et al. 2000, \apj, 537, 1054
\bibitem[Leblanc et al.(1998)]{Leb98} Leblanc, Y., Dulk, G. A., Bourgeret, J.-L. 1998, \solphys, 183, 165
\bibitem[Lewis \& Simnett(2002)]{Lew02} Lewis, D. J., \& Simnett, G. M. 2002, \mnras, 333, 969
% \bibitem[Liu et al.(2014)]{Liu14} Liu, Y. D., Luhmann, J. G., Kajdi\v{c}, P.,et al. 2014, \nat \,Communications, 5, 3481
\bibitem[Lugaz \& Kitner(2013)]{Lug13} Lugaz, N., \& Kitner, P. 2013, \solphys, 285, 281
% \bibitem[Luhmann et al.(2008)]{Luh08} Luhmann, J. G., Curtis, D. W., Schroeder, P., et al. 2008, \ssr, 136, 117
% \bibitem[Makwana et al.(2014)]{Mak14} Makwana, K. D., Zhdankin, V., Li, H., Daughton, W., Cattaneo, F. 2015, Phys. Plasmas, 22, 042902
\bibitem[Maloney, Gallagher \& McAteer(2009)]{Mal09} Maloney, S. A., Gallagher, P. T., \& McAteer, R. T. James 2009, \solphys, 256, 149 
\bibitem[Maloney \& Gallagher(2010)]{Mal10} Maloney, S. A., \& Gallagher, P. T. 2010, \apj, 724, L127
\bibitem[Manoharan et al.(2004)]{Man04} Manoharan, P. K., Gopalswamy, N., Yashiro, S.,et al. 2004, \jgr, 109, A06109
\bibitem[Manoharan(2006)]{Man06} Manoharan, P. K. 2006, \solphys, 235, 345
\bibitem[Mays et al.(2015)]{May15} Mays, M. L., Taktakishvili, A., Pulkkinen, A.,et al. 2015, arXiv:1504.04402
\bibitem[Michalek, Gopalswamy \& Yashiro(2015)]{Mic15} Michalek, G., Gopalswamy, N., \& Yashiro, S. 2015, \solphys, 290, 903
\bibitem[Mishra \& Srivastava(2013)]{Mis13} Mishra, W., \& Srivastava, N. 2013, \apj, 772, 70
\bibitem[http://omniweb.gsfc.nasa.gov()]{Omn} NASA's Space Phsyics Data Facility (SPDF), Greenbelt, MD U.S.A., NASA's Goddard Space Flight Center, http://omniweb.gsfc.nasa.gov/
\bibitem[Russell, Shinde \& Jian(2005)]{Rus05} Russell, C. T., Shinde, A. A., \& Jian, L. 2005, Adv. in Space Res, 35, 2178
\bibitem[Shaikh \& Zank(2010)]{Sha10} Shaikh, D., \& Zank, G. P. 2010, \mnras, 402, 362
\bibitem[Sheeley et al.(1997)]{She97} Sheeley, N. R., Wang, Y.-M., Hawley, S. H.,et al. 1997, \apj, 484, 472
\bibitem[Sheeley et al.(1999)]{She99} Sheeley, N. R. Jr., Walters, J. H., Wang, Y.-M.,et al. 1999, \jgr, 104, 24,739
\bibitem[Shi et al.(2015)]{Shi15} Shi, T., Wang, Y., Wan, L., et al. 2015, arXiv.1505.00884
\bibitem[Spangler(2002)]{Spa02} Spangler, S. R. 2002, \apj, 576, 997
\bibitem[St.Cyr et al.(2000)]{StC00} St. Cyr, O. C., Plunkett, S. P., Michels, D. J.,et al. 2000, \jgr, 105, 18169
\bibitem[Smith et al.(2001)]{Smi01} Smith, C. W., Mullan, D. J., Ness, N. F., et al. 2001, \jgr, 106, 18625
\bibitem[Subramanian \& Vourlidas(2007)]{Sub07} Subramanian, P., \& Vourlidas, A. 2007, \aap, 467, 685
\bibitem[Subramanian, Lara \& Borgazzi(2012)]{Sub12} Subramanian, P., Lara, A., \& Borgazzi, A. 2012, \grl, 39, L19107
\bibitem[Subramanian et al.(2014)]{Sub14} Subramanian, P., Arunbabu, K. P., Vourlidas, A.,et al. 2014, \apj, 790, 125
 \bibitem[Verma (1996)]{Ver96} Verma, M. K. 1996, JGR, 101, 27543
\bibitem[Vr\v{s}nak (2006)]{Vrs06} Vr\v{s}nak, B. 2006, Adv. in Space Res., 38, 431
\bibitem[Vr\v{s}nak et al.(2010)]{Vrs10} Vr\v{s}nak, B., \v{Z}ic, T., Falkenberg, T. V.,et al. 2010, A \& A, 512, A43
\bibitem[Vr\v{s}nak et al.(2013)]{Vrs13} Vr\v{s}nak, B., \v{Z}ic, T., Temmer, M.,et al. 2013, \solphys, 285, 295
\bibitem[Temmer et al.(2011)]{Tem11} Temmer, M., Rollett, T.,M\"{a}stl, C.,et al. 2011, \apj, 743, 101 
\bibitem[Temmer \& Nitta(2015)]{Tem15} Temmer, M., \& Nitta, N. 2015, \solphys, 290, 919
\bibitem[Thernisien, Howard \& Vourlidas(2006)]{The06}Thernisien, A. F. R., Howard, R. A., \& Vourlidas, A. 2006, \apj, 652, 763
\bibitem[Thernisien, Vourlidas \& Howard(2009)]{The09}Thernisien, A. F. R., Vourlidas, A., \&  Howard, R. A. 2009, \solphys, 256, 111
\bibitem[Thernisien(2011)]{The11}Thernisien, A. 2011, \apjs, 194, 33
\bibitem[Zhang \& Dere(2006)]{Zha06}Zhang, J., \&  Dere, K. P. 2006, \apj, 649, 1100
\bibitem[\v{Z}ic, Vr\v{s}nak \& Temmer(2015)]{Zic15} \v{Z}ic, T., Vr\v{s}nak, B., \& Temmer, M. 2015, \apjs, 218 , 7
\end{thebibliography}
\end{document}